\documentclass[10pt]{amsart}
\usepackage{amsmath}
\usepackage{latexsym}
\usepackage{amsfonts}
\usepackage{amssymb}
\usepackage{amsthm}
\usepackage{bbm}
\usepackage{dsfont}
\usepackage{graphicx}
\usepackage{color}

\newtheorem{proposition}{Proposition}
\newtheorem{theorem}{Theorem}
\newtheorem{lemma}{Lemma}
\newtheorem{corollary}{Corollary}

\theoremstyle{definition}

\newcommand{\C}{\mathbb C} 

\newcommand{\hi}{\mathcal{H}} 
\newcommand{\ip}[2]{\left\langle\,#1\,\left|\right.\,#2\,\right\rangle} 
\newcommand{\ket}[1]{\mid#1\rangle} 
\newcommand{\bra}[1]{\langle #1\mid} 
\newcommand{\kb}[2]{|#1\,\rangle\langle\,#2|} 
\newcommand{\ran}{\textrm{ran}} 
\newcommand{\rank}{\textrm{rank}} 
\newcommand{\id}{\mathbbm{1}} 
\newcommand{\nul}{\mathbbm{O}} 
\newcommand{\fii}{\varphi}

\newcommand{\sfa}{\mathsf{A}}
\newcommand{\sfb}{\mathsf{B}}

\newcommand{\sfl}{\mathsf{L}}

\newcommand{\sfi}{\mathsf{i}}

\newcommand{\Vector}[2]{ \left(
\begin{array}{c}#1\\#2 \end{array}\right) }

\newcommand{\Matrix}[9]{ \left(
\begin{array}{ccc}#1& #2 & #3\\#4 & #5 & #6 \\#7 & #8 & #9 \end{array}\right) }

\begin{document}

\title{Reduction and extremality of finite observables}

\author{Heinz-J\"urgen Schmidt}
\address{Universit\"at Osnabr\"uck, Fachbereich Physik, Osnabr\"uck, Germany}
\email{hschmidt@uni-osnabrueck.de}

\begin{abstract}
We investigate the theory of finite observables, i.~e., resolutions of the finite-dimensional identity by means of
positive operators, that have a physical interpretation in terms of measurement schemes. We focus on extremal
and rank-one observables and consider various constructions that reduce observables to simpler ones. However, these
constructions do not suffice to generate all finite extremal observables, as we show by means of counter-examples.
\end{abstract}

\maketitle


\section{Introduction}\label{sec:i}

One of the most important results of the long-lasting work on the foundations of quantum theory
is the extension of the concept of an observable \cite{L64}, \cite{K83} that has, nevertheless,
remained unacquainted to large parts of the physics community.
According to the traditional view quantum observables are mathematically described by self-adjoint
operators defined on (a dense subspace of) some Hilbert space ${\mathcal H}$.
In the more general view observables are rather described by normalised
positive operator valued (POV) measures ${\sf E}:{\mathcal A}\rightarrow{\mathcal L}({\mathcal H})$, where
${\mathcal A}$ is a $\sigma$-algebra of subsets of some {\it outcome space} $\Omega$, and ${\mathcal L}({\mathcal H})$
denotes the set of bounded linear operators of ${\mathcal H}$. A normalised POV measure ${\sf E}$ has to
satisfy ${\mathbbm O}\le {\sf E}(a)\le {\mathbbm 1}$ for all $a\in{\mathcal A}$ and  ${\sf E}(\Omega)={\mathbbm 1}_{\mathcal H}$,
see, e.~g., \cite{BLPY16} Chapter $9.3$. In the case of $\Omega={\mathbbm R}$ and ${\mathcal A}$ chosen as the $\sigma$-algebra
of Borel subsets, self-adjoint operators can be identified with the special case of projection-valued POV measures and are
called ``sharp observables", see  \cite{BLPY16}.

What are the benefits of the above-sketched generalisation of the traditional observable concept?
It turns out that attempts of physically modelling the measurement of an observable in a natural way lead to the
generalized observable concept. Then the measurement scheme of a sharp observable would appear as a kind of
idealisation that could only be approximately achieved in experimental practice.
One could counter that this kind of idealization is the normal case in theoretical physics.
However, in some cases there exist theoretical obstacles to the idealisation of sharp observables, provoking
verdicts of the form ``X is not an observable", that could be overcome by the mentioned extension of the observable concept.
A recent example is the pronounced statement ``work is not an observable" in the title of \cite{TLH07} that
has been demystified in \cite{RCP14}. Other examples are time measurements \cite{H78}, \cite{W86}, \cite{KRW09}, phase-space observables
\cite{H74}, \cite{H79}, \cite{H82}, \cite{W84},  and the localisation of photons \cite{K77}, \cite{GML15}.

In this paper we focus on the convex structure of the set of those
observables that are represented as POV measures
over a finite outcome set $\Omega$ and based on a finite-dimensional
Hilbert space ${\mathcal H}$.
An important problem is that of
finding operationally relevant characterisations of extremal POV measures.
This has been addressed in different ways by various researchers,
starting with the classic work of St\o{}rmer (1974) and with recent
contributions from \cite{P99}, \cite{APP05}, \cite{P11}, \cite{HP12}, \cite{H15}, and \cite{HP18}.
Here we will concentrate on the development of
practical schemes to construct extremal finite POV measures.
There will be a certain overlap with results of the recent paper \cite{HP18}, but in order to keep the paper self-contained we will
stick to our own proofs and only mention those cases where related results have already been published before.
Section \ref{sec:gdar} contains the general definitions and some
results characterising finite rank-one observables. The criterion for extremal observables is investigated in Section \ref{sec:eo}
and shown to be equivalent to the invertibility of a certain matrix, see Proposition \ref{p3}. We adopt the obvious strategy to construct
extremal observables from simpler ones. To this end we introduce in Section \ref{sec:dsar} the direct sum of finite observables
and identify special cases where the direct sum of extremal observables will again yield extremal ones. Generally, this would not be the case,
even for the restricted class of rank-one extremal observables, as a counter-example in Section \ref{sec:c} shows.
There are, however, other construction schemes, namely disjoint sums, tensor products and partial resolutions of observables, that preserve
extremality, see Section \ref{sec:dt}. These constructions are applied to the ``rank problem" in Section \ref{sec:r} and yield partial results
for low dimensions $h=2,\ldots,5$ but without solving the general rank problem. Summarizing, the present paper contributes to the theory of
finite observables but more work is needed to complete this theory.

\section{General definitions and results}\label{sec:gdar}

\noindent
We let ${\mathcal H}$ denote a finite-dimensional complex Hilbert space with $\dim {\mathcal H}=h$.
${\mathcal L}({\mathcal H})$ is the complex vector space of linear operators $A:{\mathcal H}\to{\mathcal H}$
equipped with the scalar product $\ip{A}{B}\equiv \mbox{Tr} A^*B$. $\ran(A)$ will denote the range
of the operator $A\in{\mathcal L}({\mathcal H})$.
An $N$-valued \emph{observable} in ${\mathcal H}$ is a sequence $\sfa=(A_1,A_2,\ldots,A_N)$ of
positive operators $A_i\in{\mathcal L}({\mathcal H})$ satisfying
\begin{equation}\label{gdar1}
\sum_{i=1}^N\;A_i\;=\;\id_{\mathcal H}\,.
\end{equation}
Sometimes it will be convenient to view an observable as a map
$\sfa:I\rightarrow {\mathcal L}({\mathcal H})$,
where $I$ is a finite set of outcomes such that $\left|I\right|=N$.
The observable with $N=1$ and  $A_1=\id_{\mathcal H}$ will be called {\em trivial}.
Positive operators $A$ bounded above by the identity (so that $\nul \le A\le \id$) are called {\em effects}.
We allow that some of the effects $A_i$ may be equal to the null operator $\nul$. Due to this choice
the set of $N$-valued observables will be closed. Moreover, any $N$-valued
observable can also be viewed as an $M$-valued observable for $M\ge N$.
The effects $A_i,\,i=1,\ldots,N,$ are also called the {\em components} of the observable $\sfa$.
$\sfa$ is an element of the linear space ${\mathcal L}({\mathcal H})^N$
and, since the constraint (\ref{gdar1}) is invariant under convex linear combinations, the set of all
$N$-valued observables in ${\mathcal H}$ will be a convex subset of ${\mathcal L}({\mathcal H})^N$,
in fact a compact convex subset. Its extremal points will be called \emph{extremal observables}.
Recall that a compact convex set is the closed convex hull of the set of its extremal points (theorem of Krein-Milman),
and hence the information about the extremal observables is sufficient to recover all observables.

An observable $\sfa$ will be called  \emph{sharp}  iff all $A_i,\;i=1,\ldots,N$ are projectors,
necessarily mutually orthogonal ones. Since projectors are extremal elements of the set of effects,
sharp observables are extremal; but extremal observables need not be sharp
for $\dim {\mathcal H}>1$.
The reason is the following: If one of the components of $\sfa$, say, $A_1$ is not a projector and hence
can be written as a non-trivial convex sum of effects, $A_1=\lambda A'_1+(1-\lambda)A''_1,\;0<\lambda <1$,
it does not follow that
also $\sfa=\lambda \sfa' +(1-\lambda)\sfa''$ holds for two different observables $\sfa'$ and $\sfa''$.
Due to the constraint (\ref{gdar1}) we would have $\nul=\sum_{i=1}^{N} (A_i-A'_i)=(1-\lambda)\sum_{i=1}^{N} (A''_i-A'_i)$
and hence the sequence of operators $(A''_i-A'_i)_{i=1,\ldots,N}$ would be linearly dependent.
If such linearly dependent sequences of operators do not exist the observable $\sfa$ may be extremal
without being sharp, see the criterion for extremality formulated in Theorem \ref{T1} below.

An observable $\sfa$ will be called a \emph{rank-one} observable iff all $A_i,\;i=1,\ldots,N$ have rank equal to one.
A rank-one observables will be written as  $\sfa=\left(\kb{e_i}{e_i}\right)_{i=1,\ldots,N}$
where the $e_i\in{\mathcal H}$ need not be normalized but will be chosen as nonzero vectors.
Its \emph{Gram matrix} $G$ is defined,  as usual,  by
\begin{equation}\label{gdar2}
G_{ij}\equiv \ip{e_i}{e_j},\;i,j=1,\ldots,N \,.
\end{equation}
It satisfies $G\ge \nul$, and $G>\nul$ (all eigenvalues positive) iff $(e_i)_{i=1,\ldots,N}$ is linearly independent. The latter
will only happen if $\sfa$ is sharp, i.~e.~if $(e_i)_{i=1,\ldots,N}$ is an orthonormal basis (ONB) in ${\mathcal H}$ (hence $N=h$).
This is a consequence of the following
\begin{proposition}\label{p1}
\phantom{}\hfill
\begin{enumerate}
\item[(i)] Two rank-one observables $\sfa,\sfb$ in ${\mathcal H}$ have the same Gram matrix iff they are
unitarily equivalent.
\item[(ii)] The Gram matrix $G$ of a rank-one observable
$\sfa=\left(\kb{e_i}{e_i}\right)_{i=1,\ldots,N}$
is an $h$-dimensional projector in $\C^N$ with non-vanishing diagonal elements.
\item[(iii)] Conversely, if $G:\C^N\longrightarrow \C^N$ is an $h$-dimensional projector
with non-vanishing diagonal elements then there exists a
rank-one observable $\sfa$ in some Hilbert space ${\mathcal H}$ with $\dim {\mathcal H}=h$ such that $G$ is its
Gram matrix.
\end{enumerate}
\end{proposition}
\noindent Obviously, the special case of a sharp rank-one observable corresponds to $G={\mathbbm 1}_N$.\\

\noindent {\em Proof:}
(i) The if-part follows immediately.\\
For the only-if-part let
$\sfa=\left(\kb{e_i}{e_i}\right)_{i=1,\ldots,N}$ and
$\sfb=\left(\kb{f_i}{f_i}\right)_{i=1,\ldots,N}$. Due to
(\ref{gdar1}) the $e_i,\;i=1,\ldots,N$ span ${\mathcal H}$. Without
loss of generality we may assume that
$\left({e_i}\right)_{i=1,\ldots,h}$ is a basis of ${\mathcal H}$,
albeit in general not an orthonormal one. It follows that the
submatrix with entries $G_{ij},\; i,j=1,\ldots,h$ is invertible and
hence also $\left({f_i}\right)_{i=1,\ldots,h}$ will form a basis of
${\mathcal H}$. Then $U e_i = f_i,\; i=1,\ldots,h$ defines an
invertible linear operator $U:{\mathcal H}\to{\mathcal H}$ which
according to $\ip{e_i}{e_j}=\ip{f_i}{f_j}=\ip{U e_i}{U e_j}$
leaves the scalar product of vectors invariant.
Hence $U$ is unitary.
Since any vector $g$ is uniquely determined by the $h$-tuples of numbers $\ip{f_i}{g}_{i=1,\dots,h}$,
the remaining vectors $f_{h+j}$ are determined by the numbers $\ip{f_i}{f_{h+j}}=\ip{e_i}{e_{h+j}}=
\ip{Ue_i}{Ue_{h+j}}=\ip{f_i}{Ue_{h+j}}$, so that $Ue_{h+j}=f_{h+j}$ for all $j=1,\ldots,N-h$.

Hence $U$ is unitary and  maps $\sfa$ onto $\sfb$.\\

\noindent
(ii) Let $\left(\ket{\nu}\right)_{\nu=1,\ldots,h}$ be an ONB in ${\mathcal H}$ and
consider the linear map $E:\C^N\to \C^h$ and the associated $h\times N$-matrix with
entries $E_{\nu i}\equiv \ip{\nu}{e_i}$.
We conclude
\begin{eqnarray}\label{gdar3a}
\left(E^*\,E\right)_{ij}&=&
\sum_{\nu=1}^h E^*_{i\nu}\,E_{\nu j} =\sum_{\nu=1}^h \overline{E_{\nu i}}\,E_{\nu j}\\ \label{gdar3b}
&=& \sum_{\nu=1}^h \ip{e_i}{\nu}\ip{\nu}{e_j}= \ip{e_i}{e_j}=G_{ij}
\;,
\end{eqnarray}
i.~e.~$E^*E=G$. On the other hand,
\begin{eqnarray}\label{gdar4a}
\left(E\,E^*\right)_{\nu\mu}&=&
\sum_{i=1}^N E_{\nu i}\,E^*_{i\mu} =\sum_{i=1}^N E_{\nu i}\, \overline{E_{\mu i}}\\ \label{gdar4b}
&=& \sum_{i=1}^N \ip{\nu}{e_i}\ip{e_i}{\mu}= \ip{\nu}{\id_{\mathcal H}\,|\,\mu}
\;,
\end{eqnarray}
i.~e.~$E\,E^*=\id_h$ (the identity in $\mathbb{C}^h$).
We conclude  $G^2=(E^*E)^2=E^*(EE^*)E=E^*\id_h E=E^*E=G$ and hence $G$ is a Hermitean idempotent,
i.~e., a projector in $\C^N$.
The rank  of $G$ equals $\mbox{Tr } G=\mbox{Tr } E^*E=\mbox{Tr } E E^*=\mbox{Tr }\id_h = h$.
Further,
$G_{ii}=\|e_i\|^2\neq 0$ for $i=1,\ldots,N$.\\

\noindent
(iii) Let $\left(\psi_\nu \right)_{\nu=1,\ldots,h}$ be an ONB in $\ran(G)$, i.~e.,
a sequence of normalised orthogonal eigenvectors
of $G$ belonging to the eigenvalue $1$. We write the $\psi_\nu$ as the columns of an $(N\times h)$-matrix
$\Psi=(\psi_1,\ldots,\psi_h)$ and consider its adjoint matrix
\begin{equation}\label{gdar5}
\Psi^*=\left(
\begin{array}{c}
\psi^*_1\\
\vdots\\
\psi^*_h
\end{array}
\right)
\equiv
(e_1,\ldots,e_N)
\;.
\end{equation}
It follows that $G=\Psi\,\Psi^*$ and $\Psi^*\,\Psi=\id_h=\sum_{i=1}^N\kb{e_i}{e_i}$.
The last equation defines a rank-one observable $\sfa=\left(\kb{e_i}{e_i}\right)_{i=1,\ldots,N}$ in
$\hi=\C^h$ with Gram matrix $G$. Note that all $e_i\neq 0$ due to the assumption $G_{ii}\neq 0$.
\qed
\\

Proposition \ref{p1} yields a constructive procedure to generate all rank-one observables up to unitary equivalence.

\section{Extremal observables}\label{sec:eo}

A family $\hi_i,\,i=1,\ldots, N$ of subspaces of $\hi$ is called {\em independent} if (i) $\hi$ is the linear span of
the $\hi_i,\,i=1,\ldots, N$ and (ii) $\sum_{i=1}^N X_i=0$ where $X_i\in \mathcal{L(H)}$ and
$\ran (X_i)\subset \hi_i$
implies $X_i=0$
for all $i=1,\ldots,N$. The second condition is equivalent to the statement that the union of any set of bases of
$\mathcal{L(H}_i)$ over $i=1,\ldots,N$ will be a linearly independent sequence of operators. The following characterization
of extremal observables is due to St{\o}rmer \cite{Stoermer74}, see also \cite{APP05}:

\begin{theorem}\label{T1}
An observable $\sfa=(A_1,\ldots,A_N)$ is extremal iff the family of subspaces
$\ran (A_i),\,i=1,\ldots,N$ is independent.
\end{theorem}

If $\sfa=(A_1,\ldots,A_N)$ is an $N$-valued observable in the Hilbert space $\hi$ and ${\mathcal S}$
a sequence of operators in $\hi$ that is the union of any set of bases of
$\mathcal{L} \left( \ran (A_i)\right)$ over $i=1,\ldots,N$,
then we will say that ${\mathcal S}$ is  {\em compatible} with $\sfa$.
Hence $\sfa$ is extremal iff there exists a linearly independent sequence ${\mathcal S}$
compatible with $\sfa$. In that case every sequence  ${\mathcal S}$  compatible with $\sfa$
will be linearly independent.

Let $\sfa=(A_1,\ldots,A_N)$ be an observable, and denote by $P_i$ the projector onto the
eigenspace ${\mathcal E}_i$ of $A_i$ belonging to the eigenvalue $1$, $\;i=1,\ldots,N$.
It follows that $\ip{\varphi}{A_j\,|\varphi}=0$ and hence $A_j\,\varphi=0$ for all $\varphi\in {\mathcal E}_i$
and $j\neq i$.\\
Hence all $P_i$ are pairwise orthogonal and
$P\equiv\sum_i P_i$ and $Q\equiv\id-P$ are projectors which commute
with all $A_i,\;i=1,\ldots,N$. Now assume $\sfa=\lambda \sfa' + (1-\lambda) \sfa'',\;0<\lambda<1$ with observables $\sfa',\;\sfa''$.
It follows that $A'_i \varphi=A''_i\varphi=\varphi$ if $A_i\varphi=\varphi$ and, consequently, all $A'_i$ and $A''_i$
commute with $P$ and $Q$, $i=1,\ldots,N$. From this we conclude that
the observable $Q\sfa Q\equiv(Q A_1 Q,\ldots,Q A_N Q)$ in $Q\hi$ is extremal iff $\sfa$ is extremal.
Let ${\mathcal V}(A_i)\equiv \ran(Q A_i Q)$, then theorem \ref{T1} can be slightly sharpened to the following:
\begin{proposition}\label{p2}
An observable $\sfa=(A_1,\ldots,A_N)$ is extremal iff the family of subspaces
$\mathcal{V}(A_i),\,i=1,\ldots,N,$ defined above, is independent in $Q\hi$.
\end{proposition}

Before elaborating the above criteria for extremality we will collect some results on independent families of subspaces.
Interestingly, this is a concept of pure linear algebra, not depending on the Hilbert space structure of the
linear spaces involved. This immediately implies Lemma \ref{L1a} below.
Nevertheless, the Hilbert space structure can be used to simplify proofs,
see, e.~g.~, the proof of Lemma \ref{L2}.

Let $h_i=\dim \hi_i,\,i=1,\ldots,N$ and $h=\dim \hi$.
\begin{lemma}\label{L1}
If the family $\hi_i,\,i=1,\ldots, N$ is independent, then
$h_i + h_j \le h$ for all $1\le i < j \le N$.
\end{lemma}
\noindent {\em Proof:}  Otherwise the subspaces $\hi_i$ and $\hi_j$ would have a non-zero intersection, which
contradicts the independence of the family $\left(\hi_1,\ldots,\hi_N\right)$. \qed

\begin{lemma}\label{L1a}
If the family $\hi_i,\,i=1,\ldots, N$ is independent and $A\in{\mathcal L}(\hi)$ is invertible
then also the family $A\,\hi_i,\,i=1,\ldots, N$ will be independent.
\end{lemma}

A family of independent subspaces $\hi_i,\,i=1,\ldots, N$ is called {\em maximal} if the
$\mathcal{L(H}_i)$ span $\mathcal{L(H)}$ or, equivalently, if $\sum_{i=1}^N h_i^2=h^2$.
The following lemma also follows from theorem $1$ in \cite{HP18}.

\begin{lemma}\label{L2}
If a family of independent subspaces $\hi_i,\,i=1,\ldots, N$ is not maximal, then there exists a one-dimensional subspace
$\hi_{N+1}$ of $\hi$ such that $\left(\hi_i\right)_{i=1,\ldots, N+1}$ is independent.
\end{lemma}
\noindent {\em Proof:} Since $\left(\hi_i\right)_{i=1,\ldots, N}$ is not maximal, there exists an $X\in \mathcal{L(H)}$ which is not
in the linear span $\sfl$ of the $\mathcal{L(H}_i),\;i=1,\ldots, N$ . If both $X+X^*\in\sfl$ and $X-X^*\in\sfl$ then
also $X\in\sfl$ in contradiction to the preceding statement. Assume first $X+X^*\notin\sfl$ and
let $X+X^*=\sum_{\mu=1}^h x_\mu P_\mu$ be the spectral decomposition of the Hermitean operator
$X+X^*$ with one-dimensional projectors $P_\mu$.

If all $P_\mu$ were elements of $\sfl$ then also $X+X^*$ would belong to $\sfl$, contrary to our assumption.
Hence at least one $P_\mu$ is not in $\sfl$ and hence linearly independent of all $\mathcal{L(H}_i)$.
The desired $\hi_{N+1}$ is given by the range of $P_\mu$.

The other case $X-X^*\notin\sfl$ is treated analogously by using the spectral decomposition
of the Hermitean operator
${\sf i}\left(X-X^*\right)=\sum_{\mu=1}^h x_\mu P_\mu$.
\qed\\

Obviously, a family of $h$ mutually orthogonal one-dimensional projections in $\hi$ yields
a corresponding family of independent one-dimensional subspaces.
According to the last lemma it is, moreover, possible to find families of independent one-dimensional subspaces
$\hi_i,\,i=1,\ldots, N$ for all values of $N$ between $N=h$ and $N=h^2$. For all these values of $N$
there exist also corresponding extremal rank-one observables.
More generally, the following holds:
\begin{lemma}\label{L3}
Let $\hi_i,\,i=1,\ldots, N$ be a family of independent subspaces. There exists an extremal observable $\sfa=(A_1,\ldots,A_N)$ such that
${\rm rank}(A_i)=\dim(\hi_i)$ for all $i=1,\ldots, N$.
\end{lemma}
\noindent Note, that is {\em not} claimed that the $\hi_i$ are the ranges of the components $A_i$ of $\sfa$
but only that they have the same dimensions.

\noindent {\em Proof:}
Define $F=\sum_{i=1}^N P_i$ where the $P_i$ are the projectors onto the  $\hi_i,\,i=1,\ldots, N$. $F$
is positive and invertible. The latter
follows since a vector $0\neq\fii\in\ker(F)$ would satisfy $0=\ip{\fii}{F\fii}=\sum_{i=1}^N \ip{\fii}{P_i\, \fii}
=\sum_{i=1}^N \left\|P_i\,\fii\right\|^2$. This implies $P_i\,\fii=0$ for all $i=1,\ldots,N$
and hence $\fii$ would be orthogonal to all $\hi_i$, in contradiction to  $\fii\neq 0$
and the condition that the $\hi_i$ span $\hi$.
Then it follows that
$\sum_{i=1}^N F^{-1/2}\,P_i\,F^{-1/2}=\id_{\mathcal H}$ and
${\rm rank}(F^{-1/2}\,P_i\,F^{-1/2})={\rm rank}(P_i) =\dim(\hi_i)$ for all $i=1,\ldots, N$.
Hence $A_i\equiv F^{-1/2}\,P_i\,F^{-1/2}\ge {\mathbbm O},\;i=1,\ldots,N,$ defines an observable
$\sfa=(A_1,\ldots,A_N)$ such that
${\rm rank}(A_i) =\dim(\hi_i)$ for all $i=1,\ldots, N$.
We have $\ran\left(A_i\right)=F^{-1/2}\,\hi_i$ and the family of subspaces $F^{-1/2}\,\hi_i,\;i=1,\ldots,N,$ is independent
since $F^{-1/2}$ is invertible, see Lemma \ref{L1a}. Hence, by Theorem \ref{T1}, $\sfa$ is extremal.
\qed\\

As an example we construct an extremal observable $\sfa=(A_1,A_2,A_3)$ in an\\ $h=3$-dimensional
Hilbert space such that $\mbox{rank}(A_1)=2$ and $\mbox{rank}(A_2)=\mbox{rank}(A_3)=1$.
We start with three projections onto independent subspaces
\begin{equation}\label{ex1}
P_1=\Matrix{1}{1}{0}{1}{1}{0}{0}{0}{0}\;,\;
P_2=\tfrac{1}{3}\;\Matrix{1}{1}{1}{1}{1}{1}{1}{1}{1}\;,\;
P_3=\tfrac{1}{3}\;\Matrix{1}{-1}{1}{-1}{1}{-1}{1}{-1}{1}\;,
\end{equation}

\noindent
and perform the construction described in the proof of lemma \ref{L3}. The result is the extremal observable
$\sfa=(A_1,\,A_2,\,A_3)$ with
\begin{eqnarray}\label{ex2a}
A_1&=&\tfrac{1}{25}\Matrix{11+4\sqrt{6}}{0}{-2-3\sqrt{6}}{0}{15}{0}{-2-3\sqrt{6}}{0}{7+2\sqrt{6}}\;,\\ \label{ex2b}
A_2&=&\Matrix{\tfrac{1}{25}(7-2\sqrt{6})}{\tfrac{-1+\sqrt{6}}{5\sqrt{5}}}{\tfrac{1}{50}(2+3\sqrt{6})}
{\tfrac{-1+\sqrt{6}}{5\sqrt{5}}}{\tfrac{1}{5}}{\tfrac{4+\sqrt{6}}{10\sqrt{5}}}
{\tfrac{1}{50}(2+3\sqrt{6})}{\tfrac{4+\sqrt{6}}{10\sqrt{5}}}{\tfrac{1}{50}(11+4\sqrt{6})}\;,\\ \label{ex2c}
A_3&=&\Matrix{\tfrac{1}{25}(7-2\sqrt{6})}{\tfrac{1}{25}(\sqrt{5}-\sqrt{30})}{\tfrac{1}{50}(2+3\sqrt{6})}
{\tfrac{1}{25}(\sqrt{5}-\sqrt{30})}{\tfrac{1}{5}}{-\tfrac{4+\sqrt{6}}{10\sqrt{5}}}
{\tfrac{1}{50}(2+3\sqrt{6})}{-\tfrac{4+\sqrt{6}}{10\sqrt{5}}}{\tfrac{1}{50}(11+4\sqrt{6})}\;.
\end{eqnarray}

Now we return to the criterion of an observable $\sfa=(A_1,\ldots,A_N)$ being extremal.
Let $R_i={{\rm ran}}(A_i)$, $d_i=\dim R_i$ for $i=1,\ldots,N$, $d=\sum_i d_i$ and
$D=\sum_i d_i^2$. Further, let $\left(\ket{i\mu} \right)_{\mu=1,\ldots,d_i}$ be a basis in $R_i$.
Recall that $\sfa$ is extremal iff the sequence of linear operators
$\left( \kb{i\mu}{i\nu} \right)_{\mu,\nu=1,\ldots,d_i,\,i=1,\ldots,N}$ is linearly independent.
Again, the Gram criterion can be invoked. It yields the condition that $A$ is extremal iff $H$ is invertible, where $H$ is
the super $D\times D$ Gram matrix with entries
\begin{equation}\label{eo1a}\begin{split}
H_{i\mu\nu,j\kappa\lambda}&=\mbox{Tr}\bigl[(\kb{i\mu}{i\nu})^*\;\kb{j\kappa}{j\lambda}\bigr]\\ 
&=\ip{i\mu}{j\kappa}\ip{j\lambda}{i\nu}=\ip{i\mu}{j\kappa}\overline{\ip{i\nu}{j\lambda}}
\;.
\end{split}
\end{equation}
This means that $H$ can be viewed as an $N\times N$-matrix of
submatrices $h_{ij},\;i,j=1,\ldots,N$ which are tensor products
of the form
\begin{eqnarray}\label{eo2a}
h_{ij}&=&a_{ij}\otimes\overline{a}_{ij}\\ \label{eo2b}
(a_{ij})_{\mu\kappa}&=&\ip{i\mu}{j\kappa} \\ \label{eo2c}
(\overline{a}_{ij})_{\nu\lambda}&=&\overline{\ip{i\nu}{j\lambda}} \mbox{  for  }
\mu,\nu=1,\ldots,d_i,\;\kappa,\lambda=1,\ldots,d_j\;.
\end{eqnarray}
We will state this result in the following
\begin{proposition}\label{p3}
An observable $\sfa=(A_1,\ldots,A_N)$ is extremal iff its super Gram matrix $H$
defined in (\ref{eo1a}) is invertible.
\end{proposition}

For the special case of rank-one observables considered above we obtain the super Gram matrix $H$ with entries
$H_{ij}=|\ip{i}{j}|^2=|G_{ij}|^2$, where (\ref{gdar2}) has been used and the choice $|i\rangle=|e_i\rangle$ has been made.
According to proposition \ref{p1},  rank-one observables can be characterized
by their ``small" Gram matrix $G$ being an $h$-dimensional projection. Now we have the additional
characterization of extremal rank-one observables by the condition  $H>0$  where $H$ is obtained by squaring
the moduli of the entries of $G$.

\section{Direct sums and reduction}\label{sec:dsar}

Let $\sfa^{(\mu)}$ be $N$-valued observables
in the Hilbert spaces $\hi_\mu,\, \mu=1,2,$ and $\hi=\hi_1\oplus\hi_2$ (orthogonal direct sum).
Defining $A_j=A_j^{(1)}+A_j^{(2)}$ for $j=1,\ldots,N$ we obtain an $N$-valued observable $A$
in $\hi$ which will be denoted as $\sfa=\sfa^{(1)}\oplus \sfa^{(2)}$. Similarly, we write
$\sfa=\bigoplus_{\mu=1}^M \sfa^{(\mu)}$ for the direct sum of more than two observables.\\

Conversely, let $\sfa=(A_1,\ldots,A_N)$ be an $N$-valued observable in the Hilbert space $\hi$ and denote by $Z(\sfa)$ the center
of the (finite-dimensional) von Neumann algebra generated by the $A_i,\, i=1,\ldots,N$. It can be easily shown that the minimal
projections $P_\mu$ in $Z(\sfa)$ sum to unity, $\sum_{\mu=1}^M P_\mu = \id$. In fact, either $\id\in Z(\sfa)$ is already minimal or it contains
a minimal projection $Q_1$. Then either $\id-Q_1$ is already minimal or it contains a minimal projection $Q_2$, and so on. After
a finite number of steps one arrives at $\sum_{\mu=1}^M P_\mu = \id$ where the $P_\mu$ are minimal projections in $Z(\sfa)$.\\
In this way one obtains observables
$\sfa^{(\mu)}=(A_1^{(\mu)},\ldots,A_N^{(\mu)})$ in the Hilbert spaces $\hi_\mu=P_\mu \hi,\;
\mu=1,\ldots,M$ by virtue of the definitions $A_i^{(\mu)}=P_\mu A_i P_\mu,\, i=1,\ldots,N$
such that $\sfa=\bigoplus_{\mu=1}^M \sfa^{(\mu)}$. This procedure will be called {\em reduction} in analogy with the reduction
of group representations. The $\sfa^{(\mu)}$ are called the {\em factor observables} of $\sfa$.
If $M=1$ then $\sfa$ is called {\em irreducible} else {\em reducible}.
In the case of an irreducible observable $\sfa$,
only multiples of $\id_\hi$ commute with all $A_i,\;i=1,\ldots,N$.
A sharp observable is reducible, except for the trivial case of $\sfa=(\id_\hi)$.\\

For a given rank-one observable $\sfa=(\alpha_1 P_1,\ldots,\alpha_N P_N)$ where the $P_i$ are one-dimensional
projectors and $\dim(\hi)>1$
there exists a simple recursive algorithm to decide whether
$\sfa$ is reducible or not. The algorithm is started by setting $Q_1=P_1$
and searching for the first $P_i$ that does not commute with $Q_1$.
If all $P_i$ would commute with $P_1=Q_1$ then $A$ is reducible.
Thus we consider the case where $Q_1$ and some $P_{i_1}$ do not commute. Let
$Q_2=Q_1\vee P_{i_1}$, i.~e.~$Q_2$ projects onto the subspace spanned by $\ran (Q_1)$ and $\ran (P_{i_1})$.
Then we proceed by looking for the first $P_i,\;i=1,\ldots,N,$ which does not commute with $Q_2$ and so on.
We thus obtain a sequence $(Q_1,Q_2,\ldots,Q_L)$ of strictly increasing projectors such that its last element
$Q_L$ commutes with all $P_i,\;i=1,\ldots,N$.
If $Q_L<\id_\hi$ the observable $\sfa$ is reducible since we have found some non-trivial projection in the center $Z(\sfa)$.
Conversely, if $\sfa$ is reducible and hence some projection ${\mathbbm O}< Q < \id_\hi$ with $Q\in Z(\sfa)$ exists
then either $P_i\le Q$ or $P_i\le \id_\hi-Q$.
Assume, without loss of generality, that $P_1=Q_1\le Q$, then $P_1$ will commute with all $P_i\le \id_\hi-Q$ and the first $P_{i_1}$
that does not commute with $Q_1$ will satisfy $P_{i_1}\le Q$. Hence $Q_2\le Q$ and so on. Finally, $Q_L\le Q<\id_\hi$.
Summarizing, the observable $\sfa$ is irreducible iff  $Q_L=\id_\hi$.\\

Applying this algorithm to the case of rank-one observables $\sfa$ in $\dim(\hi)=2$ we obtain the following alternative:
Either all $P_i,\;i=1,\ldots,N$ mutually commute in which case  $\sfa$ is reducible
or two of them do not commute, say, $[P_1,P_2]\neq\nul$. In the latter case  $\sfa$ will be irreducible.
In the former case  $\sfa$ has the form  $\sfa=(\alpha_1 P_1,\ldots,\alpha_{N_1} P_1,\beta_1 P_2,\ldots,\beta_{N_2} P_2)$
where $\sum_{i=1}^{N_1} \alpha_i= \sum_{i=1}^{N_2} \beta_i=1$ and $P_1,P_2$ are orthogonal one-dimensional
projectors. This includes the case of sharp observables with $N_1=N_2=1$.\\

If $\sfa$ is extremal, then all factor observables $\sfa^{(\mu)}$ are extremal.
This holds since ${{\rm ran}}(A^{(\mu)}_i)\subset{{\rm ran}}(A_i)$ for all $i=1,\ldots,N$ and $\mu=1,\ldots,M$.
The converse is generally not true, as will be shown below by means of a counter-example.\\
We will give an example of a direct sum which respects extremality. Let $\hi=\hi_1\oplus\hi_2$,
$\sfa=(A_1,\ldots,A_N)$ be an observable in $\hi_1$, and $\sfb=(P_1,\ldots,P_N)$ a sharp observable in $\hi_2$. Then
$\sfa\oplus \sfb=(A_1+P_1,A_2+P_2,\ldots,A_N+P_N)$.  In this case we have:
\begin{proposition}\label{p4}
The direct sum of an arbitrary observable $\sfa$ and a sharp observable $\sfb$ is extremal iff $\sfa$ is extremal.
\end{proposition}
\noindent {\em Proof:} This follows from proposition \ref{p2} since ${\mathcal V}(A_i+P_i)={\mathcal V}(A_i)$ for all $i=1,\ldots,N$.
\qed
\\

Now we return to the general question under which conditions the direct sum $A\oplus B$ of extremal observables will be extremal. As above, let
$A=(A_1,\ldots,A_N)$, $R_i={{\rm ran}}(A_i)$, $d_i=\dim R_i$ for
$i=1,\ldots,N$, $d=\sum_i d_i$, $D=\sum_i d_i^2$ and $\left(\ket{i\mu} \right)_{\mu=1,\ldots,d_i}$ be
a basis  in $R_i$. Again, $H$ will
denote the super Gram matrix with entries (\ref{eo1a}). We set $B=\underline{A}$ and denote the corresponding entities for $B$
by underlined expressions. $A\oplus\underline{A}$ will be extremal iff the union of the following
four sequences of linear operators is linearly independent:
\begin{eqnarray}\label{eo3a}
&&\left( \kb{i\mu}{i\nu} \right)_{\mu,\nu=1,\ldots,d_i,\;i=1,\ldots,N}\\ \label{eo3b}
&&\left(\underline{\ket{i\mu}}\,\underline{\bra{i\nu}}\right)_{\mu,\nu=1,\ldots,\underline{d}_i,\;i=1,\ldots,N}\\ \label{eo3c}
&&\left( \underline{\ket{i\mu}}\,\bra{i\nu} \right)_{\mu=1,\ldots,\underline{d}_i,\,\nu=1,\ldots,d_i\;i=1,\ldots,N}\\ \label{eo3d}
&&\left( \ket{i\mu}\,\underline{\bra{i\nu}} \right)_{\mu=1,\ldots,d_i,\nu=1,\ldots,\underline{d}_i\;i=1,\ldots,N}
\end{eqnarray}
These four groups of operators are mutually orthogonal. Hence the corresponding super Gram matrix of
$A\oplus\underline{A}$ can be decomposed into four blocks. The first two blocks corresponding
to (\ref{eo3a}) and (\ref{eo3b}) are copies of $H$ and $\underline{H}$ and hence invertible by assumption. The third and fourth
block corresponding to (\ref{eo3c}) and (\ref{eo3d}) are identical, hence it suffices to investigate one of them, say, the fourth one.
It is an $N\times N$-matrix $R$ with entries
\begin{eqnarray}\label{eo4a}
R_{i\mu\nu,j\kappa\lambda}&=&\mbox{Tr}\bigl[( \ket{i\mu}\,\underline{\bra{i\nu}})^*\; \ket{j\kappa}\,\underline{\bra{j\lambda}}\bigr]\\ \label{eo4b}
&=&\ip{i\mu}{j\kappa}\underline{\ip{j\lambda}{i\nu}}=\ip{i\mu}{j\kappa}\overline{\underline{\ip{i\nu}{j\lambda}}}
\;.
\end{eqnarray}
This means that $R$ can be viewed as an $N\times N$ matrix of
submatrices $r_{ij},\;i,j=1,\ldots,N$ which are tensor products
of the form
\begin{eqnarray}\label{eo5a}
r_{ij}&=&a_{ij}\otimes\overline{\underline{a}}_{ij}\\ \label{eo5b}
(a_{ij})_{\mu\kappa}&=&\ip{i\mu}{j\kappa} \\ \label{eo5c}
(\overline{\underline{a}}_{ij})_{\nu\lambda}&=&\overline{\underline{\ip{i\nu}{j\lambda}}}\\ \nonumber
 \mbox{  for  }&&
\mu=1,\ldots,d_i,\,\nu=1,\ldots,\underline{d}_i,\,\kappa=1,\ldots,d_j,\,\lambda=1,\ldots,\underline{d}_j
\;.
\end{eqnarray}
Thus the question whether $R$ is invertible, and hence whether $A\oplus\underline{A}$ is extremal,
can be decided by direct calculation in concrete cases. In general, this problem
seems to be difficult. However, if $\underline{A}$ and $A$ have the same matrices $a_{ij}$,
the comparison of (\ref{eo5a}) with (\ref{eo2a}) immediately
shows that $R=H$, and hence $R$ is invertible. Thus we obtain the following result:
\begin{proposition}\label{p5}
If $A$ and $B$ are unitarily equivalent, then $A\oplus B$ is extremal iff $A$ is extremal.
\end{proposition}

\section{Direct sum of extremal rank-one observables}\label{sec:c}

We will consider the special case where $A$ and $\underline{A}$ are rank-one observables. Then $R$
will be an $N\times N$-matrix with entries $R_{ij}=G_{ij}\,\overline{\underline{G}}_{ij}$, where $G$ and $\underline{G}$ are
the Gram matrices (\ref{gdar2}) of $A$ and $\underline{A}$. Let $(\psi_1,\ldots,\psi_{\underline{h}})$
be an ONB in ${{\rm ran}}(\underline{G})$
and consider diagonal matrices
$D_\kappa\equiv\mbox{diag}(\psi_{\kappa}):
\C^N\longrightarrow \C^N,\;\kappa=1,\ldots,\underline{h}$.

It follows that
\begin{equation}\label{c1}
R=\sum_{\kappa=1}^{\underline{h}}
D_\kappa\,G\, D_\kappa^*
\;.
\end{equation}
This shows, incidentally, that element-wise multiplication with $\overline{\underline{G}}$ is a completely positive map \cite{K83}
with Kraus operators $D_\kappa$.  If $R\ge 0$
is not invertible, then $U\equiv \mbox{Kernel }R\,\neq\{0\}$. For $u\in U$ it follows that
$0=\ip{u}{\sum_\kappa D_\kappa\,G\,D_\kappa^*\mid u}=\sum_\kappa \ip{D_\kappa^* u}{G\mid D_\kappa^* u}$.
Since $G\ge 0$, all terms in this sum must vanish, whence
\begin{equation}\label{c2}
D_\kappa^* \;U\subset\mbox{Kernel}(G) \mbox{  for all  } \kappa=1,\ldots,{\underline{h}}
\;.
\end{equation}
If $A$ and $\underline{A}$ are randomly chosen extremal rank-one observables, it turns out that (\ref{c2}) never occurs
for  $U\neq\{0\}$  and hence $A\oplus\underline{A}$ will be extremal. We conjecture that $A\oplus\underline{A}$ will be extremal
with probability $1$.
But 
extremality does not hold always; see the counter-example below.\\
However, for the special case of $N=h+1$ we can show the following:
\begin{proposition}\label{p6}
Let $A$ and $\underline{A}$ be extremal rank-one observables in the Hilbert spaces $\hi$ and $\underline{\hi}$ and $N=h+1$. Then
$A\oplus\underline{A}$ is extremal.
\end{proposition}
\noindent {\em Proof:} Due to the assumption $N=h+1$, $K\equiv\mbox{Kernel }(G)$ is one-dimensional, say, $K=\C\,v$. If all
components of $v$ would vanish except, say, $v_1\neq 0$, then $G_{11}=0$  in contradiction to proposition \ref{p1}. Hence
at least two components of $v$ must be nonzero, say, $v_1,v_2\neq 0$.\\
Assume, ad absurdum, that
 $0\neq u\in U=\mbox{Kernel }(R)$.
(\ref{c2}) implies
\begin{equation}\label{c2a}
\overline{\psi_{\kappa i}}\,u_i = z_\kappa\,v_i,\quad z_\kappa\in\C, \mbox{ for all } \kappa=1,\ldots,{\underline{h}},\mbox{ and }i=1,\ldots,N
\;.
\end{equation}
Since $v_1,v_2\neq 0$, we also have $u_1,u_2\neq 0$:
If $u_1$ would vanish, then $z_\kappa\,v_1=0$ and hence $z_\kappa=0$ for all $\kappa=1,\ldots,{\underline{h}}$.
But $u_i\neq 0$ for some $i=1,\ldots,N$,
hence $\overline{\psi_{\kappa i}}\,u_{i}=0$ and $\overline{\psi_{\kappa i}}=0$ for all $\kappa=1,\ldots,k$. This implies
$\underline{G}_{ii}=0$ which contradicts the definition of rank $1$ observables. Analogously we may show that $u_2\neq 0$.\\
Due to (\ref{c2a})
\begin{equation}\label{c3}
\overline{\psi_{\kappa 1}}=z_\kappa\,\frac{v_1}{u_1},
\quad \overline{\psi_{\kappa 2}}=z_\kappa\,\frac{v_2}{u_2},\quad
\mbox{for all }\kappa=1,\ldots,\underline{h}
\;.
\end{equation}
This means that $(\underline{e}_1,\underline{e}_2)$ is linearly dependent, using the notation introduced in (\ref{gdar5}).
Recall that $\left(\kb{\underline{e}_i}{\underline{e}_i}\right)_{i=1,\ldots,N}$ is a a rank-one observable
unitarily equivalent to $\underline{A}$.
Hence $\underline{A}$ is not extremal, in contradiction to the assumption of the proposition.
\qed
\\

According to this proposition the direct sum of extremal $3$-valued qubit observables is extremal.
The next simplest case is that of extremal $4$-valued qubit observables.
Here we have found a counter-example by
choosing the rank-one observable $\underline{A}$ as a regular tetrahedron in ${\mathcal L}(\C^2)$ and
$\mbox{kernel}(G)$ as a $2$-dimensional subspace of
$\C^4$ such that the intersection
\begin{equation}\label{cx}
u\in\bigcap_{\kappa=1}^2 \left( D_\kappa^* \right)^{-1}\;\mbox{kernel}(G)
\;.
\end{equation}
is one-dimensional. From $\mbox{kernel}(G)$ one can reconstruct the observable $A$ according to proposition \ref{p1}.
The details of the counter-example are as follows: We write
$A=\left(\kb{e_i}{e_i}\right)_{i=1,\ldots,4}$ and
$\underline{A}=\left(\kb{f_i}{f_i}\right)_{i=1,\ldots,4}$ where
\begin{eqnarray}\label{c7}
e_1&=&\Vector{\tfrac{1}{2}}{-\tfrac{i}{4}\sqrt{3}}\\
e_2&=&\Vector{\tfrac{1}{2}}{\tfrac{1}{12}\left(3+\sqrt{6}+i(3\sqrt{2}+2\sqrt{3})\right)}\\
e_3&=&\Vector{\tfrac{1}{2}}{-\tfrac{i+2\sqrt{2}}{4\sqrt{3}}}\\
e_4&=&\Vector{\tfrac{1}{2}}{\tfrac{1}{12}\left(-3+\sqrt{6}+i(-3\sqrt{2}+2\sqrt{3})\right)}\\
f_1&=&\Vector{\tfrac{1}{\sqrt{2}}}{0}\\
f_2&=&\Vector{\tfrac{1}{\sqrt{6}}}{\tfrac{1}{\sqrt{3}}}\\
f_3&=&\Vector{-\tfrac{\sqrt{3}+3i}{6\sqrt{2}}}{\tfrac{1}{\sqrt{3}}}\\
f_4&=&\Vector{\tfrac{-\sqrt{3}+3i}{6\sqrt{2}}}{
\tfrac{1}{\sqrt{3}}}
\;.
\end{eqnarray}
Both observables are extremal as can be checked by calculating the eigenvalues of the corresponding super
Gram matrices $H$ and $\underline{H}$. But $A\oplus \underline{A}$ is not extremal because the block $R$
of its super Gram matrix has the characteristic polynomial
$p(x)=\tfrac{1}{256}x(-8 + 81 x - 256 x^2 + 256 x^3)$ and hence an eigenvalue $x=0$.

\section{Disjoint sums, tensor products and partial resolutions of observables}\label{sec:dt}

There exists a variant of the direct sum of observables connected with a change of the set of outcomes.
Whereas this set is invariant under direct sums, the {\em disjoint sum} of observables will be accompanied with
a disjoint union of the respective sets of outcomes. It is defined as follows.

Let $\sfa^{(\mu)}$ be $N^{(\mu)}$-valued observables
in the Hilbert spaces $\hi_\mu,\, \mu=1,2,$ and $\hi=\hi_1\oplus\hi_2$ (orthogonal direct sum).
For notational convenience we introduce sets of indices $I^{(\mu)}$ such that
$\left|I^{(\mu)}\right|=N^{(\mu)}$ for $\mu=1,2,$ and denote the disjoint union of these sets
by $I=I^{(1)}\sqcup I^{(2)}$. Let $j\in I$ and
define $A_j=A_j^{(1)}$ for $j\in I^{(1)}$ and $A_j=A_j^{(2)}$ for $j\in I^{(2)}$.
It is easily proven that we thus obtain an $N=N^{(1)}+N^{(2)}$-valued observable $A$
in $\hi$ which will be denoted by $A=A^{(1)}\odot A^{(2)}$. Similarly, we write
$A=\bigodot_{\mu=1}^M A^{(\mu)}$ for the disjoint sum of more than two observables.
Every sharp observable can be viewed as the disjoint sum of trivial observables
$\sfa^{(\mu)}=\left(\id_{\hi_\mu}\right),\mu=1,\ldots,M$.\\

We have the following
\begin{lemma}\label{L4}
If under the preceding definitions  $\sfa^{(\mu)}$ is extremal for $\mu=1,2,$ then
$\sfa=\sfa^{(1)}\odot \sfa^{(2)}$ will be extremal.
\end{lemma}
\noindent {\em Proof:}
Let ${\mathcal S}^{(\mu)}$ be sequences of operators compatible with $\sfa^{(\mu)}$, resp., for $\mu=1,2$.
By assumption and Theorem \ref{T1} both sequences of operators ${\mathcal S}^{(\mu)}$ are linearly independent.
Since they operate in mutually orthogonal subspaces $\hi_\mu$ of $\hi$ their union ${\mathcal S}$ is also linearly independent.
Due to the definition of the disjoint sum $A=A^{(1)}\odot A^{(2)}$ the sequence ${\mathcal S}$ is compatible with $\sfa$
and hence the latter will be extremal.
 \qed\\

Another construction relevant for the present purposes is the {\em tensor product} of observables
connected with the Cartesian product of the respective sets of outcomes.
Let $A^{(\mu)}$ be $N^{(\mu)}$-valued observables
in the Hilbert spaces $\hi_\mu,\, \mu=1,2,$ and $\hi=\hi_1\otimes\hi_2$ (tensor product).
For notational convenience we use the sets of indices $I^{(\mu)}$ introduced above such that
$\left|I^{(\mu)}\right|=N^{(\mu)}$ for $\mu=1,2,$ and denote the Cartesian product of these sets
by $I=I^{(1)}\times I^{(2)}$. Let $(i,j)\in I$ and
define $A_{(i,j)}=A_i^{(1)}\otimes A_j^{(2)}$.
Recall that the tensor product of two positive operators is positive.
Moreover,
\begin{equation}\label{dt1}
  \sum_{(i,j)\in I}\,A_{(i,j)} =\sum_{i\in I^{(1)}} A_i^{(1)}\;\otimes \;\sum_{j\in I^{(2)}} A_j^{(2)}
  =\id_{\hi_1}\otimes \id_{\hi_2} =\id_{\hi}
\;.
\end{equation}
Thus we obtain an $N=N^{(1)}\times N^{(2)}$-valued observable $\sfa$
in $\hi$ which will be denoted as $\sfa=\sfa^{(1)}\otimes \sfa^{(2)}$. Similarly, we write
$\sfa=\bigotimes_{\mu=1}^M \sfa^{(\mu)}$ for the tensor product of more than two observables.\\

We have the following
\begin{lemma}\label{L5}
Under the preceding definitions, let ${\mathcal S}^{(1)}_i,\;i\in {\mathcal I}$ and
${\mathcal S}^{(2)}_j,\;j\in {\mathcal J}$  be two linearly independent sequences of
operators in $\hi_1$, resp.~$\hi_2$.
Then ${\mathcal S}_{(i,j)}={\mathcal S}^{(1)}_i\otimes {\mathcal S}^{(2)}_j$ where $(i,j)$ runs through
${\mathcal I}\times {\mathcal J}$ will also be a linearly independent sequence of operators in $\sfa^{(1)}\otimes \sfa^{(2)}$.
 \end{lemma}
\noindent {\em Proof:}
With the usual identifications we have
${\mathcal L}\left(\hi_1\otimes \hi_2\right)={\mathcal L}\left(\hi_1\right)\otimes {\mathcal L}\left(\hi_2\right)$.
Since the sequences ${\mathcal S}^{(\mu)}$ are linearly independent for $\mu=1,2,$ they can be completed to a
basis of ${\mathcal L}\left( \hi_\mu\right)$, resp., and the above defined sequence ${\mathcal S}$ can be completed to
a product basis of ${\mathcal L}\left( \hi_1\otimes \hi_2\right)$.
This product basis and its subsequence ${\mathcal S}$  are linearly independent.
\qed

Assume that the sequences ${\mathcal S}^{(\mu)}$ are compatible with the observables $\sfa^{(\mu)}$, resp.~,
then the above defined sequence ${\mathcal S}$ will be compatible with $\sfa^{(1)}\otimes \sfa^{(2)}$.
This proves the following
\begin{corollary}\label{C1}
If $\sfa^{(1)}$ and $\sfa^{(2)}$ are extremal then also $\sfa^{(1)}\otimes \sfa^{(2)}$ will be extremal.
\end{corollary}
\noindent This corollary also follows from Theorem 4.1(b) and its Corollary 4.1 of \cite{HHP14}.\\

As a third transformation of observables connected with a change of the set of outcomes we mention the {\em partial resolution}.
Let $\sfa=(A_1,\ldots,A_N)$ be an observable with a set of outcomes $I$ and
$A_n=\sum_{i=1}^{M_n} \alpha_i^{(n)}\,P_i^{(n)}$ be the spectral decomposition
of one of its components such that $\alpha_i^{(n)}>0$ and the $P_i^{(n)}$ are one-dimensional projections. Then we may
replace $A_n$ by the pair $(A'_n,A''_n)\equiv (\sum_{\stackrel{i=1}{i\neq j}}^{M_n} \alpha_i^{(n)}\,P_i^{(n)},\alpha_j^{(n)}\,P_j^{(n)})$,
where $1\le j\le M_n$ and correspondingly enlarge the outcome set $I$ to $I'$ such that $|I'|=|I|+1$.
Obviously the new sequence
$\sfa'=(A_1,\ldots,A'_n,A''_n,\ldots,A_N)$ is again an observable with $N+1$ outcomes. This transformation
$\sfa\mapsto \sfa'$ will be called an {\em elementary partial resolution}. Any transformation composed of a finite number of elementary
partial resolutions will be called a {\em partial resolution}. The maximal resolution $\hat{\sfa}$ of an observable $\sfa$ is obtained
if all its components $A_n$ are completely spectrally decomposed and hence $\hat{\sfa}$ will be a rank one observable.

Upon an elementary partial resolution  $\sfa\mapsto \sfa'$ the range $\ran A_n$ will be replaced by the two orthogonal subspaces
$\ran A'_n$ and  $\ran A''_n$ that span $\ran A_n$. Hence any sequence ${\mathcal S}'$ of operators compatible with $ \sfa'$
can be extended into a sequence ${\mathcal S}$ of operators compatible with $\sfa$. It follows that ${\mathcal S}'$ is linearly
independent if ${\mathcal S}$ is so. Hence the following holds for any partial resolution being composed of elementary ones, 
see also theorem $3$ of \cite{HP18}:

\begin{lemma}\label{L6}
If $\sfa'$ is obtained from an extremal observable $\sfa$ by means of a partial resolution then $\sfa'$ will also be extremal.
\end{lemma}

Summarizing, the disjoint sum, the tensor product and the partial resolution are constructions
that yield extremal observables if applied to extremal ones.
This principle will be applied in the next section to the ``rank problem", i.~e., the problem of which ranks may occur for the
components of extremal observables.

\section{Possible ranks of extremal observables}\label{sec:r}
\subsection{Generalities}

Let $\sfa=\left(A_i,\ldots,A_N\right)$  be an extremal observable in the Hilbert space $\hi$ with dimension $h$.
Then the corresponding numbers
$d_i\equiv\rank \left(A_i\right)=\dim\ran \left(A_i\right),\;i=1,\ldots,N$ are subject to the following constraints
\begin{eqnarray}
\label{r1a}
\sum_{i=1}^{N}d_i &\ge& h,\\
\label{r1b}
   d_i+d_j &\le & h \mbox{ for all } i,j=1,\ldots N,\\
   \label{r1c}
   \sum_{i=1}^{N}d_i^2 &\le& h^2
   \;.
\end{eqnarray}
The first condition (\ref{r1a}) holds since the subspaces $\ran \left(A_i\right)$ span $\hi$,
for (\ref{r1b}) see Lemma \ref{L1}, and (\ref{r1c}) holds since the family of subspaces $\ran \left(A_i\right)$ is
independent due to Theorem \ref{T1}.

The problem has been raised \cite{H15} whether all numbers $d_i$ subject to the conditions (\ref{r1a})--(\ref{r1c})
can be realized as the ranks of the components of suitable extremal observables. By means of Lemma \ref{L3} this ``rank problem" can be
reduced to the question whether the $d_i$ can be realized as the dimensions of a family of independent subspaces of $\hi$.
We will not solve this problem in general but rather apply the results of the previous sections to the rank problem for small $h=\dim \hi$.
A few general remarks are in order.

According to Lemma \ref{L2} it is always possible to complete a list of possible ranks to a maximal list by adding a number of $1$'s.
Hence it suffices to only consider maximal lists.
The possible ranks $d_i$ of an extremal observable will then be denoted in form of equations $h^2=\sum_i n_i\, d_i^2$, where $1^2$ may be omitted,
for example $3^2=2^2+5$ representing the maximal list $(2,1,1,1,1,1)$ of possible ranks for an extremal observable for $\dim\hi=3$.
Certain extremal observables and the corresponding rank lists can be established from general considerations.
These and the corresponding equations are called {\em standard} ones and will be defined in the following.
For example,
according to the remark preceding Lemma \ref{L3}, the extremal rank one observables with $N=h,h+1,\ldots,h^2$ yield a maximal list
$(1,\ldots,1)$ of possible ranks represented by the standard equation $h^2=h^2\times 1^2$. Further,
each partition of $h$ gives rise to a sharp, and hence extremal observable and to the corresponding list of possible ranks
that may be completed to a maximal one.
For $h=3$ we have the three partitions $3=2+1=1+1+1$ and the corresponding maximal lists represented by the standard equations $3^2=2^2+5=9$.

Another case of standard possible ranks is given if we assume that all maximal lists are known for $\dim \hi=h$. Then, according to the disjoint sum
for $\hi'=\hi\oplus \C$ and Lemma \ref{L4} the maximal lists for $\dim \hi=h$ can be completed by a suitable number of $1$'s thus obtaining
certain maximal lists for $h'=h+1$ and the corresponding standard equations. We will consider the first few dimensions $h=2,\ldots,5$ in the following subsections.
These results agree with those in section $4$ of \cite{HP18}.

\subsection{$h=2$}

For $h=\dim \hi=2$ we have the standard maximal cases mentioned above represented by $2^2=2^2$ (trivial observable) and
$2^2=4$ (maximal number of four outcomes for rank one extremal observables).

\subsection{$h=3$}

Here the standard cases due to the three partitions of $3$ are $3^2=3^2$, $3^2=2^2+5$, and $3^2=9$.
The latter two equations can also be derived from the case of $h=2$ in the manner described above.
These standard cases exhaust all possible ranks according to the conditions (\ref{r1a}) -- (\ref{r1c}).

\subsection{$h=4$}

First we consider the standard cases obtained by the five partitions $4=3+1=2+2=2+1+1=1+1+1+1$. These are represented by the equations
$4^2=3^2+7=2\times 2^2+8= 2^2+12=16$ and include the three cases derived from $h=3$. The standard cases are thus complete.
We obtain a non-standard case of possible ranks by considering the tensor product $\hi_4= \hi_2\otimes \hi_2$ with self-explaining notation
and by invoking Lemma \ref{L5} and Corollary \ref{C1}. The tensor product of the extremal observables in $\hi_2$ given by $2^2=2^2$ and $2^2=4$
thus yields the case $4^2= 4\times 2^2$ of an extremal observable in $\hi_4$ with four components consisting of rank two effects.
Alternatively, $4^2= 4\times 2^2$ can be obtained by the direct sum of two extremal observables with ranks $2^2=4\times 1^2$ according to Proposition \ref{p5}.
By a partial resolution described in Section \ref{sec:dt}  we obtain the further case $4^2= 3\times 2^2+4$,
whereas $4^2= 2\times 2^2+8$ and  $4^2= 2^2+12$ are already covered by the standard cases.

These cases exhaust all possible ranks according to the conditions (\ref{r1a}) -- (\ref{r1c}). Note, for example, that $4^2=3^2+2^2+3$ is ruled out
by (\ref{r1b}).

\subsection{$h=5$}

First we consider the standard cases obtained by the seven partitions $5=4+1=3+2=3+1+1=2+2+1=2+1+1+1=1+1+1+1+1$. These are represented by the equations
$5^2=4^2+9=3^2+2^2+12=3^2+16=2\times 2^2+17=2^2+21=25$ and include four cases derived from $h=4$.
The remaining two cases derived from $h=4$ are
$5^2=4\times 2^2+9$ and $5^2=3\times 2^2+13$ thereby completing the standard cases.

It is not possible to derive further non-standard cases by the constructions treated in this paper. However, there is one further possible rank list
given by $5^2=3^2+4\times 2^2$ together with its partial resolutions
that is compatible with (\ref{r1a}) -- (\ref{r1c}). It would be difficult to exclude this case, but fortunately it is possible
to confirm it by means of an example. Let
\begin{equation}\label{r2}
  e=\left(
\begin{array}{ccccc}
 1 & 0 & 0 & 0 & 0 \\
 0 & 1 & 0 & 0 & 0 \\
 0 & 0 & 1 & 0 & 0 \\
 0 & 0 & 0 & 1 & 0 \\
 0 & 0 & 0 & 0 & 1 \\
 1 & 1 & 1 & 0 & \sfi \\
 0 &  \sfi & 0 &  \sfi & 0 \\
 0 & 1 & 0 & 0 & 1 \\
 \sfi & 1 & 1 &  \sfi & 1 \\
 0 &  \sfi & 1 & 1 &  \sfi \\
  \sfi &  \sfi & 0 &  \sfi & 1 \\
\end{array}
\right).
\end{equation}

We consider the collection of $11$ vectors $e_i\in\C^5$ that are formed by the rows of the matrix $e$.
Then we consider the three-dimensional subspace $V_1$ spanned by $e_1, e_2, e_3$
and the four two-dimensional subspaces $V_2,\ldots, V_5$ spanned by the pairs $(e_4,e_5),\ldots , (e_{10},e_{11})$.
The linear independence of the family of subspaces $\left( V_1,\ldots, V_5\right)$ can be shown, e.~g., by calculating the determinant
$\det H=1024\neq 0$
of the super Gram matrix $H$, see Section \ref{sec:eo}.

Summarizing, we have confirmed the conjecture that the conditions  (\ref{r1a}) -- (\ref{r1c}) are not only necessary but also
sufficient for lists of possible ranks of extremal observables for the cases $h=2,\ldots,5$ but have to leave open the general
solution of the rank problem.

\section*{Acknowledgment}

An early version of this paper war co-authored by Paul Busch. Du to his untimely passing away we were not able to finish it
as a joint paper. Therefore I have to publish it under my sole responsibility but simultaneously I gratefully acknowledge his essential contributions
and, more generally, his continuous support and encouragement for research on fundamental questions of quantum theory. The community has lost
a profound scholar and a person of integrity.

\end{document}